\documentclass[conference,a4paper]{IEEEtran}
\renewcommand\IEEEkeywordsname{Index Terms}

\usepackage{amsmath,amssymb,bm}
\IfFileExists{bbm.sty}{\usepackage{bbm}}{\newcommand{\mathbbm}{\mathbb}}
\usepackage{graphicx}
\graphicspath{{fig/}}
\usepackage{booktabs}
\usepackage{cite}
\usepackage[hyphens]{url}
\usepackage[bookmarks=false]{hyperref}
\usepackage[nolist]{acronym}
\usepackage[ruled,linesnumbered]{algorithm2e}
\SetCommentSty{emph}
\usepackage{xcolor}
\usepackage{tikz}
\usetikzlibrary{calc,arrows.meta,shapes.geometric,fit,backgrounds}
\pgfdeclarelayer{background}
\pgfsetlayers{background,main}
\usepackage{pgfplots}
\pgfplotsset{compat=1.18}
\usepgfplotslibrary{groupplots}
\usepackage[caption=false,font=footnotesize]{subfig}
\definecolor{mittelblau}{RGB}{0, 126, 198}
\definecolor{violettblau}{cmyk}{0.9, 0.6, 0, 0}
\definecolor{rot}{RGB}{238, 28 35}
\definecolor{apfelgruen}{RGB}{140, 198, 62}
\definecolor{gelb}{RGB}{255, 229, 0}
\definecolor{orange}{RGB}{244, 111, 33}
\definecolor{pink}{RGB}{237, 0, 140}
\definecolor{lila}{RGB}{128, 10, 145}
\definecolor{hellgrau}{RGB}{224, 224, 224}
\definecolor{mittelgrau}{RGB}{128, 128, 128}
\definecolor{dunkelgrau}{RGB}{80,80,80}
\definecolor{anthrazit}{RGB}{19, 31, 31}
\definecolor{darkgreen}{RGB}{34,139,34}
\definecolor{aqua}{RGB}{0, 255, 255}

\definecolor{lightgray}{RGB}{211,211,211}

\definecolor{neuesgruen}{RGB}{61, 173, 65}
\definecolor{dunklereshellgrau}{RGB}{176, 176, 176}
\definecolor{neuesgelb}{RGB}{255,160,0}
\definecolor{neuescyan}{RGB}{69,185,224}

\definecolor{tollesgruen}{RGB}{0,217,171}
\definecolor{tollesmagenta}{RGB}{197,67,143}

\definecolor{tollesgelb}{RGB}{255,199,95}
\definecolor{tollesrot}{RGB}{255,111,145}

\colorlet{R12}{apfelgruen}
\colorlet{R23}{mittelblau}
\colorlet{R45}{pink}

\renewcommand{\vec}[1]{\boldsymbol{#1}}

\newcommand{\cv}{\vec{c}}

\newcommand{\sv}{\vec{s}}

\newcommand{\wv}{\vec{w}}

\newcommand{\zerov}{\vec{0}}
\newcommand{\onev}{\vec{1}}

\newcommand{\Hm}{\vec{H}}

\pgfplotsset{
  inuestyle/.style={
    width=7.2cm, height=4.6cm,
    axis x line*=bottom, axis y line*=left,
    xlabel style={font=\footnotesize}, ylabel style={font=\footnotesize},
    tick label style={font=\footnotesize},
    legend style={draw=black, fill=white, font=\scriptsize, legend cell align=left},
    grid=major, major grid style={draw=hellgrau, line width=0.35pt},
    axis line style={draw=anthrazit, line width=0.45pt},
    tick align=outside, tick style={draw=anthrazit, line width=0.45pt},
    every axis plot/.append style={line join=round, line cap=round},
  }
}

\newcommand{\Cv}{\boldsymbol{C}}

\title{Joint End-to-End Training of \\ Weighted BP Decoder Ensembles}

\author{\IEEEauthorblockN{Felix Krieg\textsuperscript{\dag}, Paul Bezner\textsuperscript{\dag}, and Stephan ten Brink}
\IEEEauthorblockA{Institute of Telecommunications, University of Stuttgart, Germany\\
\{krieg, bezner, tenbrink\}@inue.uni-stuttgart.de}
\thanks{\textsuperscript{\dag}\,P. Bezner and F. Krieg contributed equally to this work.}
\thanks{This work is supported by the German Federal Ministry of Research, Technology and Space (BMFTR) within the project Open6GHub+ (grant no. 16KIS2406).}
\thanks{LLM agents from Anthropic and OpenAI were used to write source code for this work and to assist with language editing.}}

\IEEEoverridecommandlockouts
\begin{document}
\maketitle

\begin{acronym}
\acro{AED}{automorphism ensemble decoding}
\acro{AWGN}{additive white Gaussian noise}
\acro{BCE}{binary cross-entropy}
\acro{BLER}{block error rate}
\acro{BP}{belief propagation}
\acro{BPSK}{binary phase-shift keying}
\acro{FER}{frame error rate}
\acro{GNN}{graph neural network}
\acro{LDPC}{low-density parity-check}
\acro{LLR}{log-likelihood ratio}
\acro{ML}{maximum likelihood}
\acro{WBP}{weighted belief propagation}
\acro{OSD}{ordered-statistics decoding}
\acro{PCM}{parity-check matrix}
\acro{RM}{Reed--Muller}
\acro{RNN}{recurrent neural network}
\acro{SNR}{signal-to-noise ratio}
\acro{SPA}{sum-product algorithm}

\acro{MET}{multi edge type}
\acro{CCSDS}{Consultative Committee for Space Data Systems}
\acro{PEG}{progressive edge-growth}
\acro{RBE}{row-boosted ensemble}
\acro{MBBP}{multiple basis belief propagation}
\end{acronym}

\begin{abstract}
Ensemble decoding improves decoding reliability by combining component decoders with complementary failure events. Existing approaches typically obtain this diversity from code-specific structure, such as code automorphisms, multiple parity-check representations, or targeted modifications of the message-passing graph. In this work, we instead learn the diversity of an ensemble of weighted belief propagation decoders by jointly optimizing trainable Tanner-graph edge weights. A central design question is how the training objective should distribute gradient among the ensemble members to promote complementary decoding behavior. To this end, we compare four ensemble objectives under direct gradient-based optimization and identify best-member training as the most effective approach. For a $(64,32)$ \ac{PEG} \ac{LDPC} code, the resulting ensemble outperforms the considered trained and untrained belief-propagation-based ensemble baselines. With $10$ members, it provides a gain of approximately $0.66\,\mathrm{dB}$ over a single belief propagation decoder at a target \ac{FER} of $10^{-4}$. Similar behavior is observed for a $(128,64)$ CCSDS \ac{LDPC} code, indicating that the learned complementarity is not specific to \ac{PEG}. Although the ensemble performance improves substantially, all trained members show a higher standalone \ac{FER}. The gain therefore results from learned specialization and complementary failure events rather than uniformly improved component decoders.

\end{abstract}

\begin{IEEEkeywords}
ensemble decoding, weighted belief propagation, ensemble training, short block codes
\end{IEEEkeywords}
\acresetall
\section{Introduction}
\label{sec:introduction}

Ensemble decoding improves the \ac{FER}-performance of short block codes by exploiting
diversity among several constituent decoders and selecting the most likely
candidate from their combined output list~\cite{krieg2025comparative}.
Conventional approaches create such diversity structurally, e.g., through
multiple parity-check representations~\cite{hehn2010mbbp}, code
automorphisms~\cite{geiselhart2021aed}, or decoder-side modifications of the
message-passing process~\cite{bezner2026rbe}. In contrast, learned message-passing decoders, like \ac{WBP}, improve an individual iterative decoder by optimizing trainable parameters of the decoding algorithm~\cite{nachmani2018,cammerer2022gnn}.

The two concepts have also been combined.
In~\cite{rosseel2022}, \ac{BP} \ac{RNN} decoders are trained for predefined error classes associated with absorbing sets and combined in an ensemble.
In~\cite{kwak2023boosting}, a decoder is trained on words that remain uncorrected by a preceding learned decoder.
Both approaches therefore promote specialization through prescribed or sequentially defined training subsets.

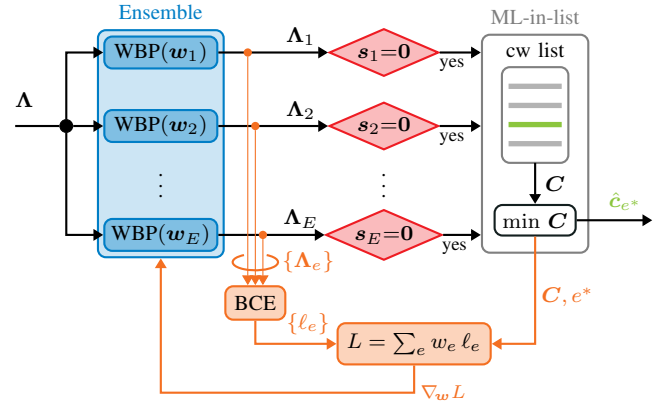
\begin{figure}[t]
\centering
\begin{tikzpicture}[
    xscale=1,
    yscale=0.72,
    font=\footnotesize,
    >={Triangle[length=1.3mm, angle'=60]},
]
\def\nbppad{0.2em}
\def\nbpwidth{4.25em}
\tikzset{nbp/.style={draw=mittelblau, fill=mittelblau!50, rounded corners=3pt,
                     align=center, thick, inner sep=2pt,
                     minimum width=\nbpwidth}}

\node (llr_in) at (1.2,0) {};
\node[circle, fill=black, draw=black, inner sep=1.7pt] (llr) at (2.0,0) {};
\node[anchor=south west] at (llr_in.north) {$\boldsymbol{\Lambda}$};

\draw[thick] (llr_in) -- (llr);

\node[nbp] (m1) at (3.25, 1.35) {{WBP}$(\wv_1)$};
\node[nbp] (m2) at (3.25, 0)    {{WBP}$(\wv_2)$};
\node      at (3.25,-0.95) {$\vdots$};
\node[nbp] (mE) at (3.25,-2.0)  {{WBP}$(\wv_E)$};

\begin{pgfonlayer}{background}
\node[draw=mittelblau, fill=mittelblau!20, thick, rounded corners=3pt,
      inner sep=\nbppad, fit=(m1)(m2)(mE)] (nbpbox) {};
\end{pgfonlayer}
\node[anchor=south, mittelblau] at (nbpbox.north) {Ensemble};
\draw[->, thick] (llr) |- (m1);
\draw[->, thick] (llr) -- (m2);
\draw[->, thick] (llr) |- (mE);

\node[shape=diamond, aspect=2.3, inner sep=1pt, align=center, draw=rot,
      fill=rot!30, thick] (v1) at (6.2,1.35) {$\sv_1{=}\zerov$};
\node[shape=diamond, aspect=2.3, inner sep=1pt, align=center, draw=rot,
      fill=rot!30, thick] (v2) at (6.2,0) {$\sv_2{=}\zerov$};
\node at (6.2,-0.95) {$\vdots$};
\node[shape=diamond, aspect=2.3, inner sep=1pt, align=center, draw=rot,
      fill=rot!30, thick] (vE) at (6.2,-2.0) {$\sv_E{=}\zerov$};
\draw[->, thick] (m1) -- (v1);
\node[anchor=south, inner sep=1.5pt] at (5.1,1.4) {$\boldsymbol{\Lambda}_1$};
\node[anchor=south, inner sep=1.5pt] at (5.1,0.05) {$\boldsymbol{\Lambda}_2$};
\node[anchor=south, inner sep=1.5pt] at (5.1,-1.95) {$\boldsymbol{\Lambda}_E$};
\draw[->, thick] (m2) -- (v2);
\draw[->, thick] (mE) -- (vE);

\node[draw=mittelgrau, fill=white, thick, rounded corners=3pt,
      minimum width=4.0em, minimum height=8.2em] (select) at (8.2,-0.325) {};
\node[anchor=south] at ($(8.2,-0.325)+(0,4 em)$) {cw list};
\draw[mittelgrau, fill=white, thick, rounded corners=3pt] ($(8.2,-0.325)+(1.25 em,4 em)$) rectangle ($(8.2,-0.325)+(-1.25 em,-1 em)$);
\draw[line width=2.0pt, mittelgrau!60] ($(8.2,-0.325)+(1.0 em,3 em)$) -- ($(8.2,-0.325)+(-1.0 em,3 em)$);
\draw[line width=2.0pt, mittelgrau!60] ($(8.2,-0.325)+(1.0 em,2 em)$) -- ($(8.2,-0.325)+(-1.0 em,2 em)$);
\draw[line width=2.0pt, apfelgruen] ($(8.2,-0.325)+(1.0 em,1 em)$) -- ($(8.2,-0.325)+(-1.0 em,1 em)$);
\draw[line width=2.0pt, mittelgrau!60] ($(8.2,-0.325)+(1.0 em,0 em)$) -- ($(8.2,-0.325)+(-1.0 em,0 em)$);
\draw[->, thick] (v1.east) -- node[pos=0.3, below, inner sep=2pt]
      {\scriptsize yes} (select.west |- v1.east);
\draw[->, thick] (v2.east) -- node[pos=0.3, below, inner sep=2pt]
      {\scriptsize yes} (select.west |- v2.east);
\draw[->, thick] (vE.east) -- node[pos=0.3, below, inner sep=2pt]
      {\scriptsize yes} (select.west |- vE.east);

\node[anchor=south, mittelgrau] at (select.north) {ML-in-list};

\node[draw=anthrazit, fill=white, rounded corners=3pt, align=center, thick, inner sep=3pt] (sel) at ($(8.2,-0.325)+(0,-4 em)$)
      {$\min\, \boldsymbol{C}$};
\draw[->, thick] ($(8.2,-0.325)+(0,-1 em)$) -- node[midway, right] {$\boldsymbol{C}$} (sel);
\draw[->, thick] (sel.east) -- ++(1.0,0)
      node[anchor=south east, text=apfelgruen]
      {$\hat{\cv}_{e^{*}}$};

\node[draw=orange, fill=orange!30, rounded corners=3pt, align=center, thick,
      inner sep=3.5pt] (loss) at (6.6,-4.0)
      {$L=\sum_e w_e\,\ell_e$};
\node[circle, fill=orange, inner sep=1.1pt, outer sep=0pt] (via1) at (4.4,1.35) {};
\node[circle, fill=orange, inner sep=1.1pt, outer sep=0pt] (via2) at (4.5,0.0) {};
\node[circle, fill=orange, inner sep=1.1pt, outer sep=0pt] (via3) at (4.6,-2.0) {};

\node[draw=orange, fill=orange!30, rounded corners=3pt, align=center, thick, inner sep=3.5pt] (bce) at (4.5,-3.25) {BCE};

\draw[->, orange] (via1) -- (via1 |- bce.north);
\draw[->, orange] (via2) -- (via2 |- bce.north);
\draw[->, orange] (via3) -- (via3 |- bce.north);

\draw[thick, orange] (4.5,-2.5) ++(60:.3 and .15) arc (60:-240:.3 and .15);

\node[text=orange] at (5.2,-2.5) {$\left\{\boldsymbol{\Lambda}_e\right\}$};

\draw[thick, orange, ->] (bce) |- node[pos=0.825, above] {$\{\ell_e\}$} (loss);

\draw[->, thick, orange] (sel.south) |- (loss.east)
      node[pos=0.28, right, inner sep=2pt] {$\boldsymbol{C},e^{*}$};

\draw[->, thick, orange] (loss.south) |- node[pos=0.5, right, inner sep=2pt, font=\scriptsize]
      {$\nabla_{\kern-.2em\wv}L$} ++(-1,-0.5) -| (nbpbox.south);
\end{tikzpicture}
\caption{Trainable ensemble model. The $E$ weighted \ac{BP} members decode the same
channel \acp{LLR}. Syndrome-valid candidates are compared using the channel
metric~\eqref{eq:metric}, and the \ac{ML}-in-list decision returns
$\hat{\cv}_{e^{*}}$. During training (\textcolor{orange}{orange}), per-frame
member losses are aggregated according to the considered objective or routing
rule.}
\label{fig:system}
\vspace{-0.75cm}
\end{figure}

In this work, we consider an ensemble of \ac{WBP} decoders, and their training without expert knowledge.
The member specialization instead emerges from the joint optimization.
This distinguishes ensemble training from single-decoder training: improving every member on average typically does not lead to the best improvement of the ensemble, because the final decision benefits from members that correct different frames.
Conversely, assigning every frame equally to every member encourages similar decoding behavior.

The central question is therefore whether the ensemble objective itself can induce member specialization under ordinary gradient-based optimization.
We compare four objectives that differ in how each frame distributes gradients across the members.
The resulting trajectories show that the objective choice materially affects the trained ensemble.
In particular, independent initialization jitter breaks member symmetry, while the best-member objective assigns each frame to its currently lowest-loss member.

The contributions of this work are summarized as follows:
  \begin{itemize}
\item We introduce a framework for joint end-to-end training of decoder
ensembles that learns complementary member behavior directly from data, without
requiring code- or decoder-specific expert knowledge.

\item We formulate four ensemble-training objectives and characterize both
their relation to the final \ac{ML}-in-list \ac{FER} and their behavior
under direct gradient-based optimization.

\item We show that the considered objectives lead to substantially different
training outcomes. Objectives that route each training sample
to a selected member outperform objectives that update the entire
ensemble simultaneously.

\item We evaluate the proposed training approach on the CCSDS\((128,64)\) and
\ac{PEG}\((64,32)\) \ac{LDPC} codes. For the \ac{PEG}\((64,32)\) code, the
resulting \(E=10\) ensemble achieves an \ac{FER} of \(10^{-4}\) at
\(0.23\,\mathrm{dB}\) lower \ac{SNR} than a specialized
ensemble baseline and provides a gain of approximately \(0.66\,\mathrm{dB}\) over a
single \ac{BP} decoder.

\item We provide evidence that the learned gains arise from member
specialization. The member failure sets have lower overlap and the ensemble \ac{FER} improves, even though the standalone \ac{FER} of every individual member degrades. 
\end{itemize}

Sec.~\ref{sec:preliminaries} introduces the decoding model, notation, reference methods, and experimental setup, while Sec.~\ref{sec:trainable} defines the candidate ensemble-training objectives. Secs.~\ref{sec:tracking} and \ref{sec:gradients} analyze the objective values and their optimization, respectively. Sec.~\ref{sec:rules} compares complete training rules, and Sec.~\ref{sec:sacrifice} analyzes the resulting member specialization.

\section{Preliminaries}
\label{sec:preliminaries}

\subsection{Code, channel, and notation}
Vectors and matrices are denoted by boldface lowercase and uppercase letters,
respectively, $\zerov$ denotes the all-zero vector, $\onev$ the all-one vector, and estimates carry a hat.
The indicator $\mathbbm{1}[\cdot]$ equals one if its argument is true and zero
otherwise. The index $j\in\{1,\ldots,n\}$ denotes a code-bit position, whereas
$e\in\{1,\ldots,E\}$ indexes a member of an ensemble of size $E$.

We consider a binary linear $(n,k)$ code
\begin{equation*}
    \mathcal{C}
    =
    \left\{
        \cv\in\mathbb{F}_2^n:
        \Hm\cv^{\mathsf T}=\zerov
    \right\}
\end{equation*}
defined by the \ac{PCM} $\Hm\in\mathbb{F}_2^{m\times n}$.
The coded bits $c_j\in\{0,1\}$ are mapped to
$x_j=1-2c_j$ and transmitted using \ac{BPSK} over an \ac{AWGN} channel,
\begin{equation*}
    y_j = x_j + z_j,
    \qquad
    z_j\sim\mathcal{N}(0,\sigma_z^2).
\end{equation*}
The corresponding channel \ac{LLR} is
\begin{equation*}
    \Lambda_j
    =
    \log
    \frac{p(y_j\mid c_j=0)}
         {p(y_j\mid c_j=1)}
    =
    \frac{2y_j}{\sigma_z^2}
\end{equation*}
and the $\Lambda_j$ are collected in
$\boldsymbol{\Lambda}=(\Lambda_1,\ldots,\Lambda_n)$. All \ac{FER} estimates are based on at least $200$ observed frame errors.

\subsection{Weighted belief-propagation decoding}

A \ac{WBP} decoder retains the message-passing structure of conventional
\ac{BP} while introducing trainable multiplicative weights into the decoding
graph~\cite{nachmani2018}. We denote the trainable parameter vector of one
decoder by $\wv$ and write
\begin{equation*}
    \boldsymbol{\Lambda}(\wv)
    =
    D_{\wv}(\boldsymbol{\Lambda})
\end{equation*}
for the posterior \acp{LLR} after iterative decoding. The hard-decision
candidate is
\begin{equation*}
    \hat{c}_j(\wv)
    =
    \mathbbm{1}\!\left[\Lambda_j(\wv)<0\right].
\end{equation*}
The decoder considered in this work uses trainable weights for
the channel contributions and exchanged messages, shared across decoding
iterations. Unit weights recover the corresponding untrained \ac{BP} decoder.

\subsection{Ensemble decoding}
An ensemble decoder applies $E$ related decoders to the same channel
observation. Member $e$ has parameters $\wv_e$, posterior
\acp{LLR}
\begin{equation*}
    \boldsymbol{\Lambda}_e
    =
    D_{\wv_e}(\boldsymbol{\Lambda})
\end{equation*}
and candidate $\hat{\cv}_e$. Its per-frame training loss is denoted by
$\ell_e$ and is specified in Sec.~\ref{sec:trainable}. Hence, diversity between
\ac{WBP} ensemble members is entirely determined by differences between their parameter vectors and, consequently, their decoding trajectories.

Only syndrome-valid candidates,
$
    \mathcal{L}
    =
    \left\{
        \hat{\cv}_e:
        \Hm\hat{\cv}_e^{\mathsf T}=\zerov
    \right\}
$
are considered for the final decision. For each valid candidate, we define the
channel metric
\begin{equation}
    C_e
    =
    \sum_{j:\,\hat{c}_{e,j}\neq b_j}
    |\Lambda_j|,
    \qquad
    b_j=\mathbbm{1}[\Lambda_j<0],
    \label{eq:metric}
\end{equation}
and collect the member metrics in
$\Cv=(C_1,\ldots,C_E)$. The final ensemble decision is
\begin{equation*}
    e^*
    =
    \operatorname*{argmin}_{e:\,\hat{\cv}_e\in\mathcal{L}} C_e,
    \qquad
    \hat{\cv}=\hat{\cv}_{e^*}.
\end{equation*}
For \ac{BPSK} transmission over the \ac{AWGN} channel, this corresponds to
\ac{ML} selection among the candidates in $\mathcal{L}$ and is the standard
selection rule used for ensemble decoding~\cite{krieg2025comparative}. If $\mathcal{L}=\varnothing$ we output a frame error.
Unless stated otherwise, \ac{FER} refers to the error rate of this
final \ac{ML}-in-list decision. When a member is evaluated individually, we explicitly refer to its \emph{standalone} \ac{FER}.

\subsection{Reference methods and experimental setup}

The main experiments use the $(64,32)$ \ac{PEG} \ac{LDPC} code referred to as Code~1 in~\cite{rosseel2022}; in the following, we denote it as the \emph{\ac{PEG} code}~\cite{hu2005peg}. To assess whether the observed behavior also occurs for another \ac{LDPC} code, we additionally consider the $(128,64)$ CCSDS \ac{LDPC} code~\cite{ccsds2310b4}, denoted as Code~2 in~\cite{rosseel2022}.
We consider ensembles of up to $E=10$ recurrent \ac{WBP} decoders. Each decoder uses one trainable weight per channel-\ac{LLR} contribution and exchanged message, shared across iterations, resulting in $256$ trainable parameters per member. 
Unless stated otherwise, training is performed with ten decoding iterations and evaluation with $25$ iterations to match the setup of~\cite{rosseel2022}. 
The training-objective analysis and the training-rule
comparison are performed at $4.0$\,dB; the resulting decoders are subsequently
evaluated over the considered \ac{SNR} range. 
The non-equivariant automorphism-groups of general \ac{LDPC} codes -- like the \ac{PEG} and CCSDS code -- are typically not known.
Hence, ensemble diversity cannot be obtained from an automorphism-based construction and must instead result from other differences between the constituent decoders.

We compare the proposed training procedure with the following reference
methods.

\emph{Single \ac{WBP}:}
A single decoder with the same architecture and parameterization as an
ensemble member is trained conventionally on channel realizations using the
bit-wise \ac{BCE} loss~\cite{nachmani2018}. This baseline distinguishes gains
due to ensemble complementarity from gains already obtainable by optimizing a single decoder.

\emph{Specialized \ac{BP}-\ac{RNN} ensemble:}
The authors of \cite{rosseel2022} obtain decoder diversity by training BP-RNN decoders specialized to error classes associated with absorbing-set structures. Their resulting ensemble is denoted by $\mathcal{D}_E$. We use the specialized BP-RNN ensemble for the \ac{PEG} code as a learned-ensemble reference and apply the same \ac{ML}-in-list selection as in~\eqref{eq:metric}. In particular, $\mathcal{D}_{10}$ denotes their ten-member ensemble used in the ensemble-size comparison.

\emph{Boosting-based training:}
The approach of \cite{kwak2023boosting} is to divide the \ac{WBP} decoder into a base decoder and a post decoder, which is specialized on words that remain uncorrected by the base decoder. Iteratively, a bigger ensemble is built using this approach. 

\emph{Row-boosted ensemble:}
The row-boosted ensemble of~\cite{bezner2026rbe} creates decoder diversity without changing the code or using trainable parameters. 
All members use the same \ac{PCM} and decoding kernel, but strengthen the outgoing messages of different subsets of parity-check rows. 
The resulting perturbed message-passing trajectories produce a set of candidate codewords that is again combined using~\eqref{eq:metric}. 
We use the optimized row-boosted ensemble as a \ac{BP}-ensemble reference with minimal learning and parameter overhead.

\emph{OSD-3:}
Finally, \ac{OSD}~\cite{fossorier1995osd} is evaluated
directly from $\boldsymbol{\Lambda}$ and serves as an approximate
\ac{ML} reference (with \ac{OSD} order $3$). It is included to indicate the remaining coding gain and is not intended as a matched-complexity baseline for the \ac{WBP} ensembles.

\section{Trainable Weighted BP Ensembles}
\label{sec:trainable}

\subsection{Ensemble training objective}

For a transmitted codeword $\cv$, the per-frame loss of member $e$ is the
bit-wise \ac{BCE} of its posterior \acp{LLR},
\begin{equation}
    \ell_e
    =
    \frac{1}{n}
    \sum_{j=1}^{n}
    \log\!\left(
        1+
        e^{-(1-2c_j)\Lambda_{e,j}}
    \right),
    \label{eq:memberloss}
\end{equation}
which decreases when the posterior \acp{LLR} move toward the signs of the
transmitted bits with increasing magnitude.

Training an individual decoder directly minimizes~\eqref{eq:memberloss}.
For an ensemble, however, the final decision depends on the collection of
candidates rather than on any individual member. We therefore express the
per-frame ensemble objective as
\begin{equation}
    L_{\bigstar}
    =
    \sum_{e=1}^{E}
    \alpha_e^\bigstar\,\ell_e,
    \label{eq:family}
\end{equation}
where the detached coefficients $\alpha_e$ determine how the gradient of a
frame is distributed among the members. The decoder parameters $\wv_e$ and
the assignment coefficients $\alpha_e$ are distinct: gradients propagate
through $\ell_e$, but not through the computation of $\alpha_e$.

\subsection{Candidate member assignments}

We consider four assignments that differ only in how member losses
contribute to~\eqref{eq:family}.

The \emph{uniform \ac{BCE}} objective assigns every frame equally to all
members,
\begin{equation}
    \alpha^{\mathrm{BCE}}_e
    =
    \frac{1}{E}.
    \label{eq:bce}
\end{equation}
This corresponds to independently training all members on the same channel
distribution and does not explicitly encourage complementary decoding
behavior.

The \emph{metric-selected objective} assigns the frame to the member whose
candidate is returned by the final ensemble selector,
\begin{equation}
    \alpha^{\mathrm{sel}}_e
    =
    \mathbbm{1}[e=e^*],
    \qquad
    e^*
    =
    \operatorname*{argmin}_{e:\,\hat{\cv}_e\in\mathcal{L}} C_e.
    \label{eq:sel}
\end{equation}
Hence, only the member responsible for the current ensemble decision receives
gradient. If $\mathcal{L}=\varnothing$, $e^*=0$.

A differentiably motivated relaxation of this assignment is the
\emph{soft metric objective},
\begin{equation}
    \boldsymbol{\alpha}^{\mathrm{soft}}
    = \operatorname{softmax}(-\widetilde{\Cv}/T),
    \label{eq:soft}
\end{equation}
where $\widetilde{C}_e=C_e$ for syndrome-valid candidates and
$\widetilde{C}_e=+\infty$ otherwise; thus, invalid candidates receive zero
weight as long as at least one member converged. We use $T=1$ in all experiments. 
If $\mathcal{L}=\varnothing$, the metric-selected objective assigns the frame to the member with the smallest $\ell_e$ and the ensemble decision itself remains a frame error.

Finally, the \emph{best-member objective} assigns the frame according to the
member losses themselves,
\begin{equation}
    \alpha^{\mathrm{best}}_e
    =
    \mathbbm{1}\!\left[
        e=
        \operatorname*{argmin}_{e'}
        \ell_{e'}
    \right].
    \label{eq:best}
\end{equation}
This assignment is analogous to the hard winner assignment of adaptive mixtures of experts~\cite{jacobs1991moe}: for each frame, the currently best member (i.e., the one that is closest to decoding this noise sample) receives the corresponding gradient.

\subsection{Ensemble-training}

We train the ensemble in a conventional manner: 
At every optimizer step, a batch of fresh channel realizations is decoded by all members, the selected per-frame losses are averaged once over the complete batch, and one joint Adam step updates all member parameter vectors.
A member that is not selected by any frame in a batch consequently receives zero gradient in that update.
Applied to the best-member assignment~\eqref{eq:best}, the same direct procedure forms the proposed training method detailed in Algorithm~\ref{alg:routed}.
An exactly symmetric initialization remains invariant for permutation-symmetric objectives because identical members process the same observations and receive identical gradients.
A hard assignment can break a tie arbitrarily, but its outcome would then depend on the arbitrary implementation of the tie-breaking rule.
To avoid this, we initialize every member from unit \ac{BP} weights with an independent Gaussian perturbation of standard deviation $\sigma=0.01$.

\section{Results}
\label{sec:results}
\subsection{Loss value versus ensemble FER}
\label{sec:tracking}
\begin{figure}[t]\centering
\input{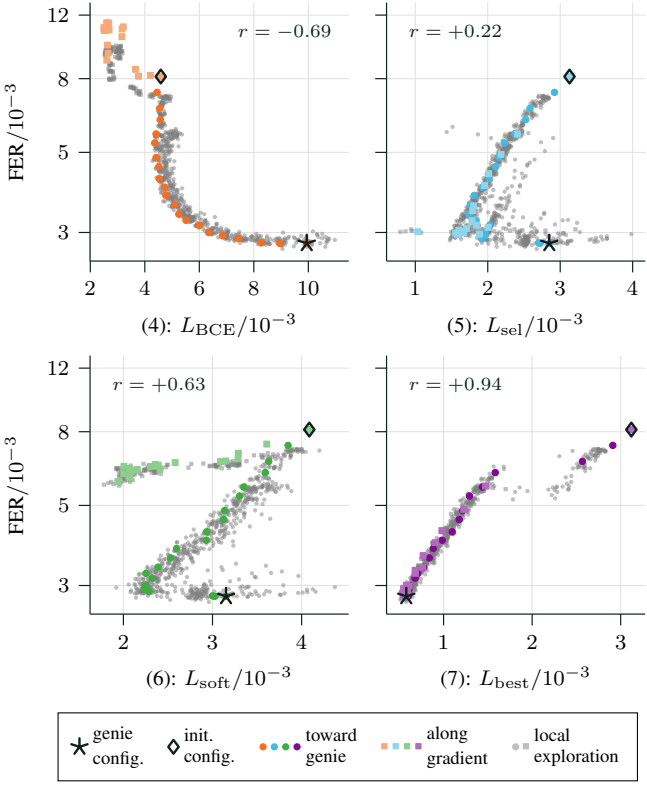}
\caption{Ensemble \ac{FER} versus loss function at $4.0\,\mathrm{dB}$ for the four losses. \ac{PEG} code with $E=10$ \ac{WBP} members. Circles denote the linear interpolation from the common initialization to the selected best-member configuration, while squares show direct Adam trajectories. Gray markers indicate local perturbations.}
\label{fig:corr}
\vspace{-0.5cm}
\end{figure}

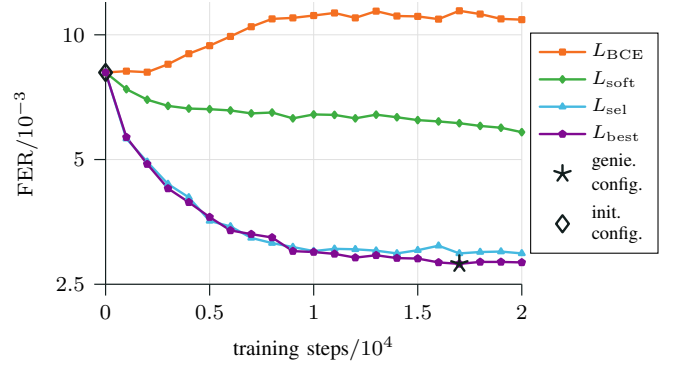
\begin{figure}
    \centering
    \begin{tikzpicture}
\begin{axis}[
  inuestyle, width=0.8\linewidth, height=0.6\linewidth,
  ymode=log, log origin=infty,
ymin=2.5e-3,
ymax=1.2e-2,
ytick={2.5e-3,5e-3,1e-2},
yticklabels={
    {$2.5$},
    {$5$},
    {$10$}
},
xmin=0,xmax=20000,
  xlabel={$\text{training steps}/10^{4}$}, ylabel={$\mathrm{FER}/10^{-3}$},
  legend style={draw=black, fill=white, font=\scriptsize,
    at={(1.02,0.5)}, anchor=west, legend columns=1},
    x tick scale label style={
    draw=none, fill=none, text opacity=0
},
]

  \addplot+[color=orange, mark=square*, line width=0.9pt, mark size=0.9pt, mark options={solid}] coordinates {(0,0.0081100463867) (1000,0.008171081543) (2000,0.0081253051758) (3000,0.0084915161133) (4000,0.0090026855469) (5000,0.0094146728516) (6000,0.0099105834961) (7000,0.010459899902) (8000,0.010925292969) (9000,0.010986328125) (10000,0.011131286621) (11000,0.011291503906) (12000,0.010986328125) (13000,0.011405944824) (14000,0.01110076904296875) (15000,0.011077880859375) (16000,0.0109100341796875) (17000,0.0114288330078125) (18000,0.01122283935546875) (19000,0.01092529296875) (20000,0.0108795166015625)};
  \addlegendentry{${L}_{\mathrm{BCE}}$}

    \addplot+[color=neuesgruen, mark=diamond*, line width=0.9pt, mark size=1.2pt, mark options={solid}] coordinates {(0,0.0081100463867) (1000,0.0073928833008) (2000,0.0069732666016) (3000,0.0067367553711) (4000,0.0066375732422) (5000,0.0066146850586) (6000,0.0065689086914) (7000,0.006462097168) (8000,0.0064926147461) (9000,0.0062866210938) (10000,0.0064239501953) (11000,0.0064086914062) (12000,0.0062866210938) (13000,0.0064163208008) (14000,0.00632476806640625) (15000,0.0062255859375) (16000,0.0061798095703125) (17000,0.0061187744140625) (18000,0.0060272216796875) (19000,0.0059661865234375) (20000,0.00582122802734375)};
  \addlegendentry{${L}_{\mathrm{soft}}$}
  
  \addplot+[color=neuescyan, mark=triangle*, line width=0.9pt, mark size=1.2pt,mark options={solid}] coordinates {(0,0.0081100463867) (1000,0.005615234375) (2000,0.0049362182617) (3000,0.0043640136719) (4000,0.0040512084961) (5000,0.0035552978516) (6000,0.0034484863281) (7000,0.0032348632812) (8000,0.0031433105469) (9000,0.0030746459961) (10000,0.0030059814453) (11000,0.003044128418) (12000,0.0030364990234) (13000,0.0030136108398) (14000,0.00296783447265625) (15000,0.003021240234375) (16000,0.0030975341796875) (17000,0.00296783447265625) (18000,0.00299072265625) (19000,0.00299835205078125) (20000,0.00296783447265625)};
  \addlegendentry{${L}_{\mathrm{sel}}$}

  \addplot+[color=lila, mark=pentagon*, line width=0.9pt, mark size=1.2pt, mark options={solid}] coordinates {(0,0.0081100463867) (1000,0.0056610107422) (2000,0.0048751831055) (3000,0.0042572021484) (4000,0.0039443969727) (5000,0.0036315917969) (6000,0.0033721923828) (7000,0.003303527832) (8000,0.0032424926758) (9000,0.0030059814453) (10000,0.0029907226562) (11000,0.0029602050781) (12000,0.0028991699219) (13000,0.0029373168945) (14000,0.00289154052734375) (15000,0.0028839111328125) (16000,0.0028228759765625) (17000,0.00279998779296875) (18000,0.00283050537109375) (19000,0.00283050537109375) (20000,0.0028228759765625)};
  \addlegendentry{${L}_{\mathrm{best}}$}

  \addplot+[only marks,
    mark=star,
    mark size=3.2pt,
    line width=0.9pt,
    anthrazit] 
    coordinates {(17000,0.00279998779296875)};
  \addlegendentry{\shortstack[l]{genie.\\config.}}
  
    \addplot+[only marks,
    mark=diamond,
    mark size=3.2pt,
    line width=0.9pt,
    anthrazit] 
    coordinates {(0,0.0081100463867)};
  \addlegendentry{\shortstack[l]{init.\\config.}}
\end{axis}
\end{tikzpicture}
\vspace{-0.4cm}
    \caption{Validation \ac{FER} versus training steps for direct Adam optimization of the four candidate objectives from a common jittered initialization.}
    \label{fig:fer_training_loss}
    \vspace{-0.4cm}
\end{figure}
Training an ensemble requires specifying how the member-wise training losses are combined. For conventional \ac{WBP} decoding, the bit-wise \ac{BCE} is the standard training objective~\cite{nachmani2018,cammerer2022gnn}. Applying the same objective to an ensemble by averaging the member losses, however, does not explicitly account for the final ensemble decision. We therefore compare four ensemble objectives derived from the same per-member \ac{BCE}.

We first investigate how the scalar objective values relate to the resulting ensemble \ac{FER}. Figure~\ref{fig:corr} evaluates \ac{BP}-weight configurations along the linear interpolation between the common jittered initialization and the genie-selected best-member checkpoint (circles), together with local perturbations around this path (gray circles). This interpolation provides a controlled path through parameter space toward a known good ensemble and is not generated by an optimizer. For comparison, the squares show configurations obtained by direct Adam optimization of the respective objective, with local perturbations shown as gray squares. The common initialization is marked by $\diamond$, whereas $\star$ denotes the genie configuration.

The four objectives differ substantially in how their scalar values relate to the ensemble \ac{FER}. The uniform \ac{BCE} is anticorrelated with \ac{FER} ($r=-0.69$), since averaging the member losses does not distinguish between improvements on the same frames and improvements on complementary error events. Consequently, a low average member loss can coincide with a comparatively high ensemble \ac{FER}.

The strongest positive correlation is obtained for the best-member objective~\eqref{eq:best} with $r=+0.94$, followed by the soft metric objective~\eqref{eq:soft} with $r=+0.63$ and the metric-selected objective~\eqref{eq:sel} with $r=+0.22$. The weaker correlation of the metric-selected objective results from its hard member selection: small changes in the decoder weights can switch the selected candidate and thus change the objective value abruptly, while the ensemble \ac{FER} changes only marginally. Thus, the correlation characterizes how well the scalar objective orders the sampled configurations, but does not directly characterize the gradients encountered during optimization.

\subsection{Optimization with candidate objectives}
\label{sec:gradients}
To separate the relation between the scalar objective value and the resulting optimization behavior, we train the same ensemble with each of the four objectives for $20{,}000$ Adam updates. All runs start from the same jittered unit-weight initialization, use a learning rate of $10^{-3}$, and process batches of $128$ channel realizations at $4.0\,\mathrm{dB}$ (drawn fresh each time; no sample is used twice during training). 
Training is performed with ten decoding iterations, while each checkpoint is evaluated with $25$ iterations on a fixed validation set. The square trajectories in Fig.~\ref{fig:corr} show one checkpoint every $1000$ updates, and Fig.~\ref{fig:fer_training_loss} resolves the $20{,}000$ updates in more detail.

The four objectives lead to clearly different optimization behavior. The validation \ac{FER} of the uniform-\ac{BCE} objective increases, whereas the soft metric objective reduces it slightly. The two hard-decision based losses perform best. 
Despite its low Pearson coefficient, $L_\mathrm{sel}$ yields strong ensemble performance. This discrepancy is caused by the final part of the interpolation path toward the genie configuration in Fig.~\ref{fig:corr}: several configurations achieve a slightly lower ensemble \ac{FER} while exhibiting a substantially larger loss, which weakens the overall linear correlation.

Since each optimizer step uses newly generated channel realizations, the degradation of performance for longer training cannot be attributed to overfitting to a finite training set. Instead, it may result from the mismatch between training and inference depth: weights optimized for ten decoding iterations do not necessarily preserve their performance when the decoder is unfolded to $25$ iterations.\footnote{Rosseel \emph{et al.}~\cite{rosseel2022} deliberately train with ten iterations and evaluate with $25$ iterations, reporting a performance gain for this configuration. We observed that this mismatch can instead degrade performance for some training objectives. Nevertheless, we retain the same setup to ensure comparability with~\cite{rosseel2022} and across the considered baselines.}

More importantly, the best-member trajectory demonstrates that direct joint optimization alone can substantially reduce the ensemble \ac{FER}. The similarly strong performance of the metric-selected objective, despite its small Pearson coefficient, further shows that the scalar loss--\ac{FER} correlation is useful for characterizing an objective, but does not by itself predict its optimization performance.

\subsection{Training rules under matched data exposure}
  \label{sec:rules}
\begin{figure}[t]
  \centering

\subfloat[\ac{PEG}\((64,32)\).]{%
  \begin{minipage}[t][4cm][t]{0.485\columnwidth}
    \centering
    \begin{tikzpicture}
\begin{axis}[
  inuestyle,
  width=1.1\linewidth,
  height=1.4\linewidth,
  ymode=log,
  log origin=infty,
  ymin=2.5e-3,
  ymax=1.2e-2,
  ytick={2.5e-3,5e-3,1e-2},
  yticklabels={
    {$2.5$},
    {$5$},
    {$10$}
  },
  x filter/.code={
    \pgfmathparse{\pgfmathresult*1e6}
  },
  xmin=0,
  xmax=6e6,
  xtick={0,1e6,2e6,3e6,4e6,5e6,6e6},
  scaled x ticks=base 10:-6,
  xlabel={$\text{training samples}/10^{6}$},
  ylabel={$\mathrm{FER}/10^{-3}$},
  legend to name=exposurelegend,
  legend columns=3,
  legend cell align={left},
  legend style={
    draw=black,
    fill=white,
    font=\scriptsize,
    inner xsep=2pt,
    inner ysep=2pt,
  },
  x tick scale label style={
    draw=none,
    fill=none,
    text opacity=0
  },
]

\addplot+[
  color=anthrazit,
  dotted,
  mark=none,
  line width=0.9pt,
  no marks
] coordinates {
  (0,0.00981140137)
  (6.5,0.00981140137)
};
\addlegendentry{\shortstack[l]{single BP\\(untrained)}}

\addplot+[
  color=dunkelgrau,
  dashed,
  mark=none,
  line width=0.85pt,
  no marks
] coordinates {
  (0,0.00695800781)
  (6.5,0.00695800781)
};
\addlegendentry{\shortstack[l]{$E=10$ RBE\\(untrained)~\cite{bezner2026rbe}}}

\addplot+[
  color=red,
  solid,
  mark=diamond,
  mark size=1.3pt,
  line width=0.9pt,
  mark options={solid}
] coordinates {
  (0,0.0081100463867)
  (0.128,0.0056610107422)
  (0.256,0.0048751831055)
  (0.512,0.0039443969727)
  (0.768,0.0033721923828)
  (1.024,0.0032424926758)
  (1.28,0.0029907226562)
  (1.536,0.0028991699219)
  (1.792,0.0028915405273)
  (2.048,0.0028228759766)
  (2.304,0.0028305053711)
  (2.56,0.0028228759766)
};
\addlegendentry{\shortstack[l]{$E{=}10$ $L_\mathrm{best}$ [prop.]}}

\addplot+[
  color=neuesgruen,
  dotted,
  mark=triangle*,
  mark size=1.3pt,
  line width=0.9pt,
  mark options={solid}
] coordinates {
  (0,0.00981140137)(0.128,0.00955200195) (0.256,0.00938415527)
  (0.512,0.00915527344)
  (1.024,0.00868225098)
  (1.536,0.00866699219)
  (2.048,0.00912475586)
  (2.56,0.00938415527)
  (3.072,0.00970458984)
  (3.584,0.0100402832)
  (4.096,0.0103607178)
  (4.608,0.0104675293)
  (5.12,0.0107421875)
  (5.632,0.0108337402)
  (6.144,0.0108642578)
};
\addlegendentry{\shortstack[l]{single WBP~\cite{nachmani2018}}}

\addplot+[
  color=orange,
  dashed,
  mark=triangle*,
  mark size=1.3pt,
  line width=0.9pt,
  mark options={solid}
] coordinates {
  (0,0.00981140137) (0.128,0.00965881348) (0.256,0.00933837891)
  (0.512,0.0076751709)
  (1.024,0.00698852539)
  (1.536,0.00662231445)
  (2.048,0.00569152832)
  (2.56,0.00549316406)
  (3.072,0.00523376465)
  (3.584,0.00495910645)
  (4.096,0.00480651855)
  (4.608,0.00463867188)
  (5.12,0.00454711914)
  (5.632,0.00448608398)
  (6.144,0.00434875488)
};
\addlegendentry{$E{=}10$ residual~\cite{kwak2023boosting}}

\addplot+[
  color=mittelblau,
  solid,
  mark=*,
  mark size=1.2pt,
  line width=0.9pt,
  mark options={solid}
] coordinates {
  (0,0.00981140137) (0.128,0.00592041016) (0.256,0.00547790527) 
  (0.512,0.00492858887)
  (1.024,0.00468444824)
  (1.536,0.00448608398)
  (2.048,0.00444030762)
  (2.56,0.00457763672)
  (3.072,0.0046081543)
  (3.584,0.00477600098)
  (4.096,0.00495910645)
  (4.608,0.00500488281)
  (5.12,0.00505065918)
  (5.632,0.00509643555)
};
\addlegendentry{\shortstack[l]{$E{=}10$ specialists~\cite{rosseel2022}}}

\end{axis}
\end{tikzpicture}
  \end{minipage}%
}\hfill
\subfloat[CCSDS\((128,64)\).]{%
  \begin{minipage}[t][4cm][t]{0.485\columnwidth}
    \centering
    \begin{tikzpicture}
  \begin{axis}[
    inuestyle,
    width=1.1\linewidth,
    height=1.4\linewidth,
    ymode=log,
    log origin=infty,
    ymin=0.9e-3,
    ymax=9e-3,
    ytick={1e-3,2e-3,4e-3,8e-3},
    yticklabels={{$1$},{$2$},{$4$},{$8$}},
    x filter/.code={
      \pgfmathparse{\pgfmathresult*1e6}
    },
    xmin=0,
    xmax=5e6,
xtick={0,1e6,2e6,3e6,4e6,5e6},
scaled x ticks=base 10:-6,
    x tick scale label style={
      draw=none,
      fill=none,
      text opacity=0
    },
    xlabel={$\text{training samples}/10^{6}$},
  ]

\addplot+[color=red, solid, mark=diamond, mark size=1.3pt,
    line width=0.9pt, mark options={solid}, mark repeat=2]
    coordinates {(0,0.00292205810547) (0.1000,0.00220489501953) (0.2000,0.00189971923828) (0.3000,0.0018310546875) (0.4000,0.00178527832031) (0.5000,0.00168609619141) (0.6000,0.00162506103516) (0.7000,0.00160980224609) (0.8000,0.00154113769531) (0.9000,0.00149536132812) (1.0000,0.00149536132812) (1.1000,0.00138854980469) (1.2000,0.00135803222656) (1.3000,0.00132751464844) (1.4000,0.00128936767578) (1.5000,0.00121307373047) (1.6000,0.00113677978516) (1.7000,0.00121307373047) (1.8000,0.00121307373047) (1.9000,0.00122833251953) (2.0000,0.00118255615234) (2.1000,0.00118255615234) (2.2000,0.00114440917969) (2.3000,0.00120544433594) (2.4000,0.00119018554688) (2.5000,0.00116729736328) (2.6000,0.00111389160156) (2.7000,0.00113677978516) (2.8000,0.00113677978516) (2.9000,0.00108337402344) (3.0000,0.00107574462891) (3.1000,0.00106811523438) (3.2000,0.00105285644531) (3.3000,0.00104522705078) (3.4000,0.00102233886719) (3.5000,0.00105285644531) (3.6000,0.00100708007812) (3.7000,0.00106811523438) (3.8000,0.00104522705078) (3.9000,0.00102996826172) (4.0000,0.00102233886719) (4.1000,0.00103759765625) (4.2000,0.00102996826172) (4.3000,0.00104522705078) (4.4000,0.00102996826172) (4.5000,0.00100708007812) (4.6000,0.000961303710938) (4.7000,0.000968933105469) (4.8000,0.000991821289062) (4.9000,0.000999450683594) (5.0000,0.0009765625)};
  \addplot+[color=mittelblau, solid, mark=*, mark size=1.2pt,
    line width=0.9pt, mark options={solid}, mark repeat=2]
    coordinates {(0,0.00495147705078) (0.1000,0.00193023681641) (0.2000,0.00171661376953) (0.3000,0.00172424316406) (0.4000,0.00164794921875) (0.5000,0.00157165527344) (0.6000,0.00152587890625) (0.7000,0.00149536132812) (0.8000,0.00148773193359) (0.9000,0.00150299072266) (1.0000,0.00148010253906) (1.1000,0.00154113769531) (1.2000,0.00160217285156) (1.3000,0.001708984375) (1.4000,0.00181579589844) (1.5000,0.00183868408203) (1.6000,0.00187683105469)};
  \addplot+[color=orange, dashed, mark=triangle*, mark size=1.2pt,
    line width=0.9pt, mark options={solid}, mark repeat=2]
    coordinates {(0,0.00495147705078) (0.1000,0.00470733642578) (0.2000,0.00457763671875) (0.3000,0.00340270996094) (0.4000,0.00334167480469) (0.5000,0.00331878662109) (0.6000,0.00324249267578) (0.7000,0.00325012207031) (0.8000,0.0028076171875) (0.9000,0.00277709960938) (1.0000,0.00270843505859) (1.1000,0.002685546875) (1.2000,0.00266265869141) (1.3000,0.00248718261719) (1.4000,0.00248718261719) (1.5000,0.00242614746094) (1.6000,0.00241088867188) (1.7000,0.00240325927734) (1.8000,0.00238037109375) (1.9000,0.00230407714844) (2.0000,0.00226593017578) (2.1000,0.00217437744141) (2.2000,0.00218200683594) (2.3000,0.002197265625) (2.4000,0.00210571289062) (2.5000,0.00208282470703) (2.6000,0.00204467773438) (2.7000,0.00204467773438) (2.8000,0.00201416015625) (2.9000,0.00197601318359) (3.0000,0.00196075439453) (3.1000,0.001953125) (3.2000,0.00198364257812) (3.3000,0.001953125) (3.4000,0.00194549560547) (3.5000,0.00187683105469) (3.6000,0.00187683105469) (3.7000,0.00187683105469) (3.8000,0.00184631347656) (3.9000,0.00184631347656) (4.0000,0.00182342529297) (4.1000,0.00181579589844) (4.2000,0.00178527832031) (4.3000,0.00179290771484) (4.4000,0.00180053710938) (4.5000,0.00177001953125) (4.6000,0.00176239013672) (4.7000,0.00176239013672) (4.8000,0.00173187255859) (4.9000,0.00172424316406) (5.0000,0.00173950195312)};
  \addplot+[color=neuesgruen, dashed, mark=triangle*, mark size=1.2pt,
    line width=0.9pt, mark options={solid}, mark repeat=2]
    coordinates {(0,0.00495147705078) (0.1000,0.0048828125) (0.2000,0.00480651855469) (0.3000,0.00452423095703) (0.4000,0.00444793701172) (0.5000,0.00426483154297) (0.6000,0.00432586669922) (0.7000,0.00418853759766) (0.8000,0.004150390625) (0.9000,0.00400543212891) (1.0000,0.0040283203125) (1.1000,0.00394439697266) (1.2000,0.00393676757812) (1.3000,0.00392150878906) (1.4000,0.00383758544922) (1.5000,0.00373840332031) (1.6000,0.00370788574219) (1.7000,0.00363159179688) (1.8000,0.00362396240234) (1.9000,0.00351715087891) (2.0000,0.00347137451172) (2.1000,0.00341796875) (2.2000,0.00347900390625) (2.3000,0.00341796875) (2.4000,0.00343322753906) (2.5000,0.00333404541016) (2.6000,0.00331115722656) (2.7000,0.00324249267578) (2.8000,0.00321197509766) (2.9000,0.00325012207031) (3.0000,0.00321197509766) (3.1000,0.00319671630859) (3.2000,0.00319671630859) (3.3000,0.00321960449219) (3.4000,0.00319671630859) (3.5000,0.00321960449219) (3.6000,0.00318145751953) (3.7000,0.00312805175781) (3.8000,0.00316619873047) (3.9000,0.00314331054688) (4.0000,0.00314331054688) (4.1000,0.00315856933594) (4.2000,0.00310516357422) (4.3000,0.00311279296875) (4.4000,0.00314331054688) (4.5000,0.00318145751953) (4.6000,0.00322723388672) (4.7000,0.00322723388672) (4.8000,0.00320434570312)};

\addplot+[color=anthrazit, dotted, mark=none, mark size=1.2pt,
    line width=0.9pt, mark options={solid}, mark repeat=2]
    coordinates {(0,0.00495147705078) (6,0.00495147705078)};
\addplot+[color=dunkelgrau, dashed, mark=none, mark size=1.2pt,
line width=0.9pt, mark options={solid}, mark repeat=2]
coordinates {(0,0.00245666503906) (6,0.00245666503906)};

  \end{axis}
  \end{tikzpicture}
  \end{minipage}%
}

  \par\medskip
  \pgfplotslegendfromname{exposurelegend}

  \caption{Validation \ac{FER} versus cumulative training-sample
  exposure at an SNR of \(4\,\mathrm{dB}\) for \(E=10\) ensembles
  using minibatches of \(128\) samples.}
  \label{fig:exposure}
\vspace{-0.4cm}
\end{figure}
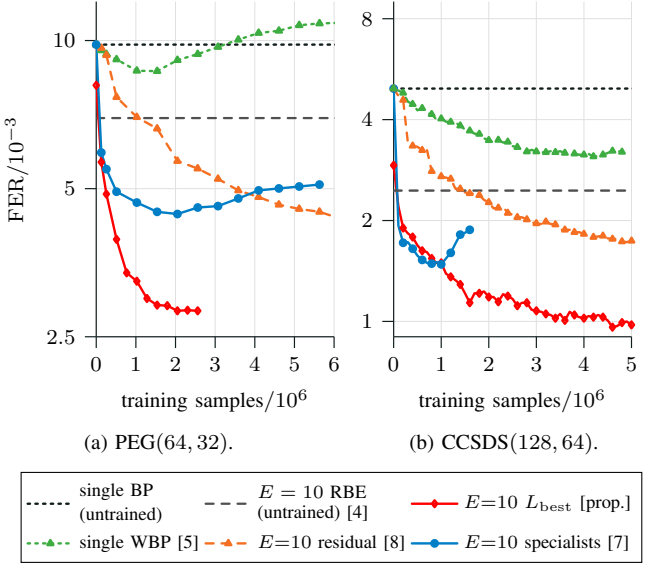

\begin{figure}[t]
  \centering
  \begin{tikzpicture}
\begin{axis}[
  inuestyle,
  width=0.70\linewidth, height=0.70\linewidth,
  ymode=log, log origin=infty,
  xmin=2.5, xmax=5, ymin=3e-4, ymax=2e-1,xtick={2.5,3,3.5,4,4.5,5},
  xlabel={$E_b/N_0$ (dB)}, ylabel={FER},
    grid=both,
  minor grid style={dashed, draw=hellgrau},
  legend style={draw=black, fill=white, font=\scriptsize,
    at={(1.02,0.5)}, anchor=west, legend columns=1}
]
  \addplot+[color=anthrazit, dotted, mark=*, mark size=1.1pt, line width=0.9pt, mark options={solid}] coordinates {(2.5,0.136108398) (2.75,0.0973899148) (3,0.0667317708) (3.25,0.0434782609) (3.5,0.0264199747) (3.75,0.0170056573) (4,0.00929576946) (4.25,0.00581287202) (4.5,0.00318416633) (4.75,0.00160515445) (5,0.000765331113)};
  \addlegendentry{\shortstack[l]{single BP\\(untrained)}}
  
    \addplot+[color=neuesgruen, dotted, mark=triangle*, mark size=1.3pt, line width=0.9pt, mark options={solid}] coordinates {(2.5,0.145368304) (2.75,0.100195312) (3,0.0688802083) (3.25,0.0448774858) (3.5,0.0253405449) (3.75,0.0157335069) (4,0.00851732337) (4.25,0.00511289267) (4.5,0.0025390625) (4.75,0.00121857488) (5,0.000555812464)};
  \addlegendentry{\shortstack[l]{single WBP~\cite{nachmani2018}}}
  
  \addplot+[color=dunkelgrau, dashed, mark=square*, mark size=1.1pt, line width=0.9pt, mark options={solid}] coordinates {(2.5,0.119032118) (2.75,0.0843098958) (3,0.0556640625) (3.25,0.0355398996) (3.5,0.0209025931) (3.75,0.0127587256) (4,0.00675511853) (4.25,0.00385626845) (4.5,0.0018356438) (4.75,0.000891024179)};
  \addlegendentry{\shortstack[l]{$E{=}10$ RBE\\(untrained)~\cite{bezner2026rbe}}}

  \addplot+[color=mittelblau, solid, mark=*, mark size=1.2pt, line width=0.9pt, mark options={solid}] coordinates {(2.5,0.108940972) (2.75,0.0716378348) (3,0.0463423295) (3.25,0.0272894965) (3.5,0.0152740479) (3.75,0.00876290458) (4,0.004272994) (4.25,0.00231644328) (4.5,0.00102365042) (4.75,0.000442283741)};
  \addlegendentry{\shortstack[l]{$E{=}10$\\specialists~\cite{rosseel2022}}}
  
  \addplot+[color=orange, dashed, mark=triangle*, mark options={rotate=180}, mark size=1.3pt, line width=0.9pt, mark options={solid}] coordinates {(2.5,0.117296007) (2.75,0.0782001202) (3,0.0495605469) (3.25,0.0289522059) (3.5,0.016389974) (3.75,0.00889559659) (4,0.0042917352) (4.25,0.00232979911) (4.5,0.000974613273) (4.75,0.000423911129)};
  \addlegendentry{\shortstack[l]{$E{=}10$\\residual~\cite{kwak2023boosting}}}

  \addplot+[color=red, solid, mark=diamond, mark size=1.3pt, line width=1.1pt, line width=0.9pt, mark options={solid}] coordinates {(2.5,0.0849609375) (2.75,0.0549587674) (3,0.0323791504) (3.25,0.0186993634) (3.5,0.0102335612) (3.75,0.0055027651) (4,0.0025507609) (4.25,0.00135257964) (4.5,0.000590424728)};
  \addlegendentry{\shortstack[l]{$E{=}10$, $L_\mathrm{best}$\\ \hspace{.1pt}[prop.] }}
  
  \addplot+[color=black, solid, mark=none, line width=0.9pt] coordinates {(2.5,0.015531994) (2.75,0.00902200401) (3,0.00539537293) (3.25,0.00283061594) (3.5,0.00146191991) (3.75,0.000774755048)};
  \addlegendentry{OSD-3 ($\approx$ML)}
\end{axis}
\end{tikzpicture}
\vspace{-0.7cm}
\caption{\ac{FER} versus $E_b/N_0$ for checkpoints selected on
the \ac{PEG} code after training as shown in Fig.~\ref{fig:exposure} (a).}
\label{fig:fer_ebno}
\end{figure}
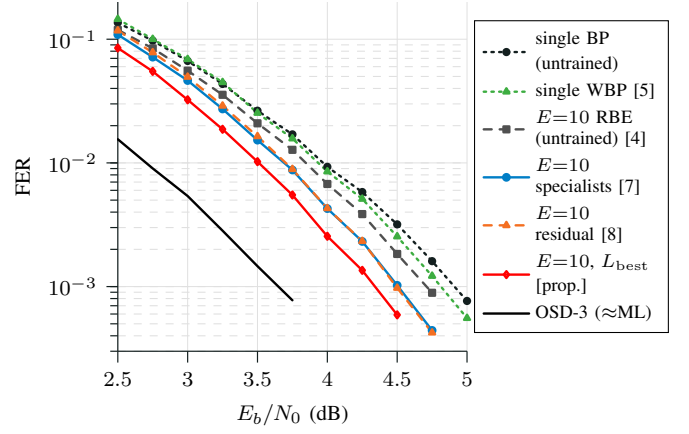

\SetAlgoSkip{} 

\begin{algorithm}[t]
\caption{Joint \ac{WBP} ensemble training via $L_\mathrm{best}$}
\label{alg:routed}
\KwIn{ensemble size $E$; jitter $\sigma$; batch size $B$;
      number of updates $U$; learning rate $\eta$}
\KwOut{trained parameters $\wv_1,\ldots,\wv_E$}

\tcp{initialize diverse members around conventional BP}
\For{$e=1,\ldots,E$}{
    $\boldsymbol{\xi}_e
    \sim \mathcal{N}(\zerov,\sigma^2\mathbf{I})$\;
    $\wv_e \gets \mathbf{1}+\boldsymbol{\xi}_e$\;
}

\For{$u=1,\ldots,U$}{
    draw $B$ independent channel realizations
    $\{\boldsymbol{\Lambda}^{(b)}\}_{b=1}^{B}$\;

    \For{$b=1,\ldots,B$}{
        \For{$e=1,\ldots,E$}{
            $\boldsymbol{\Lambda}_e^{(b)}
            \gets D_{\wv_e}(\boldsymbol{\Lambda}^{(b)})$\;
            $\ell_e^{(b)}
            \gets \ell(\boldsymbol{\Lambda}_e^{(b)},\cv^{(b)})$\;
        }
        $a_b \gets
        \operatorname*{argmin}_{e}
        \operatorname{stopgrad}\!\left(\ell_e^{(b)}\right)$
        \tcp*[r]{no gradient through assignment; ties resolved randomly}
    }

    $L_{\mathrm{best}}
    \gets
    \frac{1}{B}
    \sum_{b=1}^{B}
    \ell_{a_b}^{(b)}$\;

    update $\wv_1,\ldots,\wv_E$ jointly using Adam on
    $\nabla L_{\mathrm{best}}$ with learning rate $\eta$\;
}
\end{algorithm}

To compare the learned methods under a common training budget, we use \emph{optimizer sample exposure}. Each frame entering a shared minibatch is counted once, independent of the number of ensemble members that receive gradient from it, while samples not used in an optimizer update are excluded. This measure therefore reflects the number of channel realizations contributing to gradient updates without scaling joint ensemble updates by the ensemble size.

For the proposed method in Algorithm~\ref{alg:routed}, we use $E=10$, $B=128$, $\sigma=0.01$, and Adam with a learning rate of $10^{-3}$ at $4.0\,\mathrm{dB}$. All members remain trainable for $20{,}000$ joint updates, corresponding to $2.56\cdot10^6$ training samples. As shown in Fig.~\ref{fig:exposure}, direct best-member training reduces the validation \ac{FER} rapidly and achieves the lowest value among the considered learned methods for both codes.

Figure~\ref{fig:fer_ebno} evaluates the checkpoints with minimal \ac{FER} in Fig.~\ref{fig:exposure} on fresh channel realizations over the considered \ac{SNR} range. The directly trained best-member ensemble achieves the lowest \ac{FER} among the considered \ac{BP}-based methods throughout this range, demonstrating that the gain is not limited to the training \ac{SNR}.

\begin{figure}[t]
\centering
\begin{tikzpicture}
\begin{axis}[
  inuestyle,
  width=0.7\linewidth, height=0.6\linewidth,
  xmin=1, xmax=10, ymin=5, ymax=5.7,
  xtick={1,3,...,19},
  xlabel={ensemble size $E$}, ylabel={SNR for FER $=10^{-4}$ in dB},
  legend style={draw=black, fill=white, font=\scriptsize,
    at={(1.05,0.5)}, anchor=west, legend columns=1},
]
  \addplot+[
      color=dunkelgrau,
      dashed,
      mark=square*,
      mark size=1.1pt,
      line width=0.9pt,
      mark options={solid}
  ] coordinates {
      (1,5.67458056)
      (2,5.58368258)
      (3,5.54302449)
      (4,5.50933929)
      (5,5.49193361)
      (6,5.46989463)
      (7,5.44429110)
      (8,5.42848397)
      (9,5.42150195)
      (10,5.41183151)
  };
  \addlegendentry{\shortstack[l]{RBE\\(untrained)~\cite{bezner2026rbe}}}
  
\addplot+[color=neuesgruen, dotted, mark=triangle*, mark size=1.4pt, line width=1.1pt, mark options={solid}]
  coordinates {
      (1,5.69296556)
      (2,5.59714975)
      (3,5.52036456)
      (4,5.46144117)
      (5,5.43322531)
      (6,5.41206822)
      (7,5.37937283)
      (8,5.31836351)
      (9,5.29868557)
      (10,5.29045257)
  };
  \addlegendentry{\shortstack[l]{MBBP\\(untrained)~\cite{hehn2010mbbp}}}
  \addplot+[color=mittelblau, solid, mark=*, mark size=1.2pt, line width=0.9pt, mark options={solid}] coordinates {(1,5.6) (2,5.512) (3,5.446) (4,5.423) (5,5.36) (6,5.34) (7,5.323) (8,5.286) (9,5.271) (10,5.25)};
  \addlegendentry{specialists \cite{rosseel2022}}
  \label{plt:rosseel}
  \addplot+[color=red, solid, mark=diamond, mark size=1.4pt, line width=1.1pt] coordinates {(1,5.54418875) (2,5.40273326) (3,5.24737795) (4,5.20249825) (5,5.17360073) (6,5.14360946) (7,5.08028890) (8,5.04710457) (9,5.02214304) (10,5.01653435)};
  \addlegendentry{{$L_\mathrm{best}$ [prop.]}}
\end{axis}
\end{tikzpicture}
\vspace{-0.2cm}
\caption{Required \ac{SNR} for \ac{FER} $10^{-4}$ versus ensemble size $E$ for the \ac{PEG} code. The specialists curve reproduces Fig.~6 of~\cite{rosseel2022}.}
\label{fig:rosseel6}
\vspace{-0.3cm}
\end{figure}
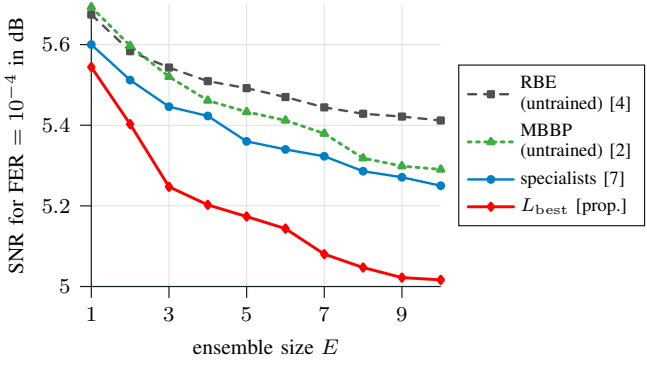

Figure~\ref{fig:rosseel6} shows the required \ac{SNR} for different ensemble sizes $E$. For the proposed method, the \ac{SNR} decreases rapidly with increasing $E$ and remains below that of the untrained specialist ensembles for all considered ensemble sizes. At $E=10$, it reaches $5.02\,\mathrm{dB}$, compared with $5.25\,\mathrm{dB}$ for $\mathcal{D}_{10}$~\cite{rosseel2022}.

\subsection{Member specialization and ensemble complementarity}
\label{sec:sacrifice}
\begin{figure}[b]\centering
\begin{tikzpicture}
\begin{axis}[
  inuestyle,
  width=8.6cm, height=6.0cm,
  xlabel={ensemble member},
  ylabel={FER},
  ymode=log,
  xmin=0.3, xmax=10.3,
  ymin=1.6e-3, ymax=1.2e-1,
  xtick={1,2,...,10},
  legend style={at={(0.03,0.97)}, anchor=north west},
  legend cell align=left,
  >={Triangle[length=1.3mm, angle'=60]},
  grid=both,
  minor grid style={dashed, draw=hellgrau},
]
\addplot[mittelblau, only marks, mark=*, mark size=1.8pt]
  coordinates {(1,1.028824e-02) (2,1.033783e-02) (3,1.033783e-02) (4,1.031876e-02) (5,1.025772e-02) (6,1.035690e-02) (7,1.039124e-02) (8,1.028061e-02) (9,1.025009e-02) (10,1.029587e-02)};
\addlegendentry{initial member}
\addplot[rot, only marks, mark=triangle*, mark size=2.2pt]
  coordinates {(1,2.446747e-02) (2,2.460098e-02) (3,2.368927e-02) (4,2.750778e-02) (5,2.302170e-02) (6,2.819443e-02) (7,2.688217e-02) (8,2.893066e-02) (9,2.090073e-02) (10,2.026749e-02)};
\addlegendentry{trained member}
\addplot[mittelblau, dashed, line width=0.7pt, forget plot]
  coordinates {(0.3,8.090973e-03) (13.6,8.090973e-03)};
\node[font=\scriptsize, text=mittelblau, anchor=north west]
  at (axis cs:0.55,7.767334e-03) {ensemble at initialization (gain from jitter)};
\addplot[rot, dashed, line width=0.7pt, forget plot]
  coordinates {(0.3,2.799988e-03) (13.6,2.799988e-03)};
\node[font=\scriptsize, text=rot, anchor=south west]
  at (axis cs:0.55,3.086548e-03) {trained ensemble};
\draw[->, black,
  line width=0.8pt] (axis cs:8.5,7.767334e-03)
  -- node[right, font=\scriptsize, text=black, inner sep=2pt, align=left]
  {ensemble \\ gains} (axis cs:8.5,2.967834e-03);

\draw[->, black,
  line width=0.8pt] (axis cs:8.5,1e-02)
  -- node[right, font=\scriptsize, text=black, inner sep=2pt, align=left, pos=0.35]
  {members \\ lose} (axis cs:8.5,3.5e-02);
\end{axis}
\end{tikzpicture}
\vspace{-0.2cm}
\caption{Standalone and ensemble \ac{FER} before and after joint training for the \ac{PEG} code with $E=10$ \ac{WBP} members at $4.0\,\mathrm{dB}$. Markers denote the standalone member \acp{FER}, dashed lines the complete ensemble \ac{FER}.}
\label{fig:sacrifice}
\end{figure}
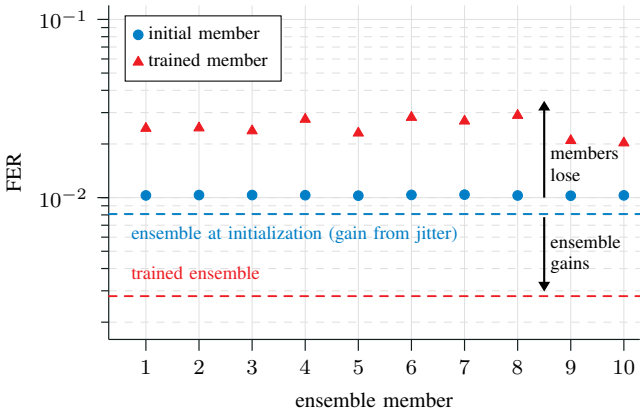

To assess whether the ensemble gain results from stronger individual decoders or increased complementarity, we evaluate all members before and after the proposed training. 
Fig.~\ref{fig:sacrifice} shows the \ac{FER} of the ensemble and the respective standalone \ac{FER} of the members.
Although the ensemble \ac{FER} improves from $8.09\cdot10^{-3}$ to $2.80\cdot10^{-3}$, the respective standalone \ac{FER} increases, indicating complementarity.

Let $\mathcal{F}e$ denote the set of frames not decoded by member $e$. We quantify pairwise failure overlap by the Jaccard index
\begin{equation}
J_{e,e'} =
\frac{\lvert\mathcal{F}e \cap \mathcal{F}{e'}\rvert}
{\lvert\mathcal{F}e \cup \mathcal{F}{e'}\rvert}.
\label{eq:jaccard}
\end{equation}
For channel realizations at $4.0\,\mathrm{dB}$, the mean Jaccard index over all $45$ member pairs decreases from $0.885$ to $0.315$.

Thus, the ensemble gain is accompanied by substantially less overlap between the member failure sets. The best-member assignment~\eqref{eq:best} promotes this specialization by updating the member with the lowest loss on each frame rather than improving all members jointly. Consequently, the trained ensemble benefits from complementary frame-specific behavior despite degraded standalone member performance.

\section{Conclusion}
\label{sec:conclusion}
Training complementary \ac{WBP} decoder ensembles requires an objective that promotes member specialization. For the considered \ac{PEG} code, the four investigated objectives differ substantially in both their scalar relation to the ensemble \ac{FER} and their behavior under direct optimization. The best-member objective achieves the strongest correlation and the lowest trained \ac{FER}, while the competitive performance of the metric-selected objective despite its weak correlation shows that correlation of scalar loss and \ac{FER} alone does not predict trainability.
Using joint best-member training from independently jittered unit-weight initializations, the resulting $E=10$ ensemble achieves the lowest \ac{FER} among the considered \ac{BP}-based methods throughout the evaluated \ac{SNR} range and requires $0.23,\mathrm{dB}$ less \ac{SNR} than the published specialized ensemble at \ac{FER} $10^{-4}$. At $4.0,\mathrm{dB}$, the ensemble \ac{FER} improves by a factor of $2.9$ although the standalone \ac{FER} of every member increases. Together with the reduced overlap of the member failure sets, this shows that the ensemble gain results from learned specialization and complementary error events rather than uniformly improved constituent decoders.
{\global\widowpenalty=10000}

\bibliographystyle{IEEEtran}
\bibliography{references}

\end{document}